\documentclass{webofc}
\usepackage[varg]{txfonts}   
\begin{document}

\def\planck{{\it Planck}} 
\def\aj{AJ}%
\def\araa{ARA\&A}%
\def\apj{ApJ}%
\def\apjl{ApJ}%
\def\apjs{ApJS}%
\def\aap{Astron. Astrophys.}%
 \def\aapr{A\&A~Rev.}%
\def\aaps{A\&AS}%
\def\mnras{MNRAS}
\def\ssr{SSRv}
\def\nat{Nature}
\def\jcap{JCAP}

\def\moo{MOO\,J1142$+$1527}
\def\xmm{XMM-{\it Newton}}
\def\planck{{\it Planck}} 
\def\chandra{{\it Chandra}}
\def \rosat {\hbox{\it ROSAT}}
\newcommand{\excpres}{{\gwpfont EXCPRES}}
\title{Mapping the gas thermodynamic properties of the massive cluster merger \moo\ at z = 1.2}

\author{\firstname{F.} \lastname{Ruppin} \inst{\ref{MIT}}\fnsep\thanks{\email{ruppin@mit.edu}}
\and  \firstname{R.} \lastname{Adam} \inst{\ref{LLR}}
\and  \firstname{P.} \lastname{Ade} \inst{\ref{Cardiff}}
\and  \firstname{P.} \lastname{Andr\'e} \inst{\ref{CEA1}}
\and  \firstname{A.} \lastname{Andrianasolo} \inst{\ref{IPAG}}
\and  \firstname{M.} \lastname{Arnaud} \inst{\ref{CEA1}}
\and  \firstname{H.} \lastname{Aussel} \inst{\ref{CEA1}}
\and  \firstname{I.} \lastname{Bartalucci} \inst{\ref{CEA1}}
\and  \firstname{M. W.} \lastname{Bautz} \inst{\ref{MIT}}
\and  \firstname{A.} \lastname{Beelen} \inst{\ref{IAS}}
\and  \firstname{A.} \lastname{Beno\^it} \inst{\ref{Neel}}
\and  \firstname{A.} \lastname{Bideaud} \inst{\ref{Neel}}
\and  \firstname{O.} \lastname{Bourrion} \inst{\ref{LPSC}}
\and  \firstname{M.} \lastname{Brodwin} \inst{\ref{UMis}}
\and  \firstname{M.} \lastname{Calvo} \inst{\ref{Neel}}
\and  \firstname{A.} \lastname{Catalano} \inst{\ref{LPSC}}
\and  \firstname{B.} \lastname{Comis} \inst{\ref{LPSC}}
\and  \firstname{B.} \lastname{Decker} \inst{\ref{UMis}}
\and  \firstname{M.} \lastname{De~Petris} \inst{\ref{Roma}}
\and  \firstname{F.-X.} \lastname{D\'esert} \inst{\ref{IPAG}}
\and  \firstname{S.} \lastname{Doyle} \inst{\ref{Cardiff}}
\and  \firstname{E.~F.~C.} \lastname{Driessen} \inst{\ref{IRAMF}}
\and  \firstname{P. R. M.} \lastname{Eisenhardt} \inst{\ref{JPL}}
\and  \firstname{A.} \lastname{Gomez} \inst{\ref{CAB}}
\and  \firstname{A. H.} \lastname{Gonzalez} \inst{\ref{UFlo}}
\and  \firstname{J.} \lastname{Goupy} \inst{\ref{Neel}}
\and  \firstname{F.} \lastname{K\'eruzor\'e} \inst{\ref{LPSC}}
\and  \firstname{C.} \lastname{Kramer} \inst{\ref{IRAME}}
\and  \firstname{B.} \lastname{Ladjelate} \inst{\ref{IRAME}}
\and  \firstname{G.} \lastname{Lagache} \inst{\ref{LAM}}
\and  \firstname{S.} \lastname{Leclercq} \inst{\ref{IRAMF}}
\and  \firstname{J.-F.} \lastname{Lestrade} \inst{\ref{LERMA}}
\and  \firstname{J.F.} \lastname{Mac\'ias-P\'erez} \inst{\ref{LPSC}}
\and  \firstname{P.} \lastname{Mauskopf} \inst{\ref{Cardiff},\ref{Arizona}}
\and  \firstname{F.} \lastname{Mayet} \inst{\ref{LPSC}}
\and  \firstname{M.} \lastname{McDonald} \inst{\ref{MIT}}
\and  \firstname{A.} \lastname{Monfardini} \inst{\ref{Neel}}
\and  \firstname{E.} \lastname{Moravec} \inst{\ref{UFlo}}
\and  \firstname{L.} \lastname{Perotto} \inst{\ref{LPSC}}
\and  \firstname{G.} \lastname{Pisano} \inst{\ref{Cardiff}}
\and  \firstname{E.} \lastname{Pointecouteau} \inst{\ref{IRAP}}
\and  \firstname{N.} \lastname{Ponthieu} \inst{\ref{IPAG}}
\and  \firstname{G. W.} \lastname{Pratt} \inst{\ref{CEA1}}
\and  \firstname{V.} \lastname{Rev\'eret} \inst{\ref{CEA1}}
\and  \firstname{A.} \lastname{Ritacco} \inst{\ref{IRAME}}
\and  \firstname{C.} \lastname{Romero} \inst{\ref{IRAMF}}
\and  \firstname{H.} \lastname{Roussel} \inst{\ref{IAP}}
\and  \firstname{K.} \lastname{Schuster} \inst{\ref{IRAMF}}
\and  \firstname{S.} \lastname{Shu} \inst{\ref{IRAMF}}
\and  \firstname{A.} \lastname{Sievers} \inst{\ref{IRAME}}
\and  \firstname{S. A.} \lastname{Stanford} \inst{\ref{UCal}}
\and  \firstname{D.} \lastname{Stern} \inst{\ref{JPL}}
\and  \firstname{C.} \lastname{Tucker} \inst{\ref{Cardiff}}
\and  \firstname{R.} \lastname{Zylka} \inst{\ref{IRAMF}}}

\institute{
Kavli Institute for Astrophysics and Space Research, Massachusetts Institute of Technology, Cambridge, MA 02139, USA
\label{MIT}
\and
Laboratoire Leprince-Ringuet, \'Ecole Polytechnique, CNRS/IN2P3, 91128 Palaiseau, France
\label{LLR}
\and
Astronomy Instrumentation Group, University of Cardiff, UK
  \label{Cardiff}  
\and
AIM, CEA, CNRS, Universit\'e Paris-Saclay, Universit\'e Paris Diderot, Sorbonne Paris Cit\'e, F-91191 Gif-sur-Yvette, France
\label{CEA1}
\and
Univ. Grenoble Alpes, CNRS, IPAG, 38000 Grenoble, France
  \label{IPAG}
\and
Institut d'Astrophysique Spatiale (IAS), CNRS and Universit\'e Paris Sud, Orsay, France
  \label{IAS}
\and
Institut N\'eel, CNRS and Universit\'e Grenoble Alpes, France
  \label{Neel}
\and
Univ. Grenoble Alpes, CNRS, Grenoble INP, LPSC-IN2P3, 53, avenue des Martyrs, 38000 Grenoble, France
  \label{LPSC}
  \and
Department of Physics and Astronomy, University of Missouri, 5110 Rockhill Road, Kansas City, MO 64110, USA
  \label{UMis}
\and
Dipartimento di Fisica, Sapienza Universit\`a di Roma, Piazzale Aldo Moro 5, I-00185 Roma, Italy
  \label{Roma}
\and
Institut de RadioAstronomie Millim\'etrique (IRAM), Grenoble, France
  \label{IRAMF}
  \and
Jet Propulsion Laboratory, California Institute of Technology, Pasadena, CA 91109, USA
\label{JPL}
\and
Centro de Astrobiolog\'ia (CSIC-INTA), Torrej\'on de Ardoz, 28850 Madrid, Spain
\label{CAB}
\and
Department of Astronomy, University of Florida, 211 Bryant Space Center, Gainesville, FL 32611, USA
\label{UFlo}
\and
Institut de RadioAstronomie Millim\'etrique (IRAM), Granada, Spain
  \label{IRAME}
\and
Aix Marseille Universit\'e, CNRS, LAM (Laboratoire d'Astrophysique de Marseille) UMR 7326, 13388, Marseille, France
  \label{LAM}
\and 
LERMA, Observatoire de Paris, PSL Research University, CNRS, Sorbonne Universit\'es, UPMC Univ., 75014, Paris, France
  \label{LERMA}
\and
School of Earth and Space Exploration and Department of Physics, Arizona State University, Tempe, AZ 85287
  \label{Arizona}
  \and 
IRAP, Universit\'e de Toulouse, CNRS, CNES, UPS, (Toulouse), France
  \label{IRAP}
\and 
Institut d'Astrophysique de Paris, CNRS (UMR7095), 98 bis boulevard Arago, F-75014, Paris, France\label{IAP}\and Department of Physics, University of California, One Shields Avenue, Davis, CA 95616, USA
  \label{UCal}
}      

\abstract{
We present the results of the analysis of the very massive cluster \moo\ at a redshift $z = 1.2$ based on high angular resolution NIKA2 Sunyaev-Zel’dovich (SZ) and \chandra\ X-ray data. This multi-wavelength analysis enables us to estimate the shape of the temperature profile with unprecedented precision at this redshift and to obtain a map of the gas entropy distribution averaged along the line of sight. The comparison between the cluster morphological properties observed in the NIKA2 and \chandra\ maps together with the analysis of the entropy map allows us to conclude that \moo\ is an on-going merger hosting a cool-core at the position of the X-ray peak. This work demonstrates how the addition of spatially-resolved SZ observations to low signal-to-noise X-ray data can bring valuable insights on the intracluster medium thermodynamic properties at $z>1$.}

\maketitle

\section{Introduction}\label{sec:intro}

The analysis of the evolution of the gas thermodynamic properties with the mass and redshift of galaxy clusters enables the characterization of the processes behind cluster growth and the underlying cosmology in which these processes take place. Extensive analyses of the intracluster medium (ICM) astrophysical processes have been conducted primarily in X-ray in the direction of $z<1$ clusters \emph{e.g.} \citep{pra10}. However, the most active part of cluster formation is expected at redshifts $1 < z < 2$ \emph{e.g.} \citep{poo07}. Therefore, the ICM thermodynamic properties may have significantly evolved since $z{\sim}2$. This may induce significant modifications of the cosmological constraints obtained from the analysis of cluster abundance \citep{rup19}. It is thus essential to extend our knowledge of the ICM properties at redshifts higher than 1.\\
The \emph{Massive and Distant Clusters of WISE Survey} (MaDCoWS) is an Infra-Red (IR) survey currently conducted in order to detect the most massive galaxy clusters at $z \gtrsim  1$ \citep{gon18}. In this paper, we present a multi-wavelength analysis of the MaDCoWS cluster \moo\ at $z=1.19$ combining high angular resolution SZ data obtained  by the NIKA2 camera at the Institut de Radioastronomie Millimétrique (IRAM) 30-m telescope \cite{ada18} and X-ray data from \chandra. This joint analysis allows us to estimate the radial distributions of all thermodynamic properties and to produce maps of their average values along the line of sight.

\section{Observations of \moo}\label{sec:observations}

\begin{figure*}
\centering
\includegraphics[height=4.7cm]{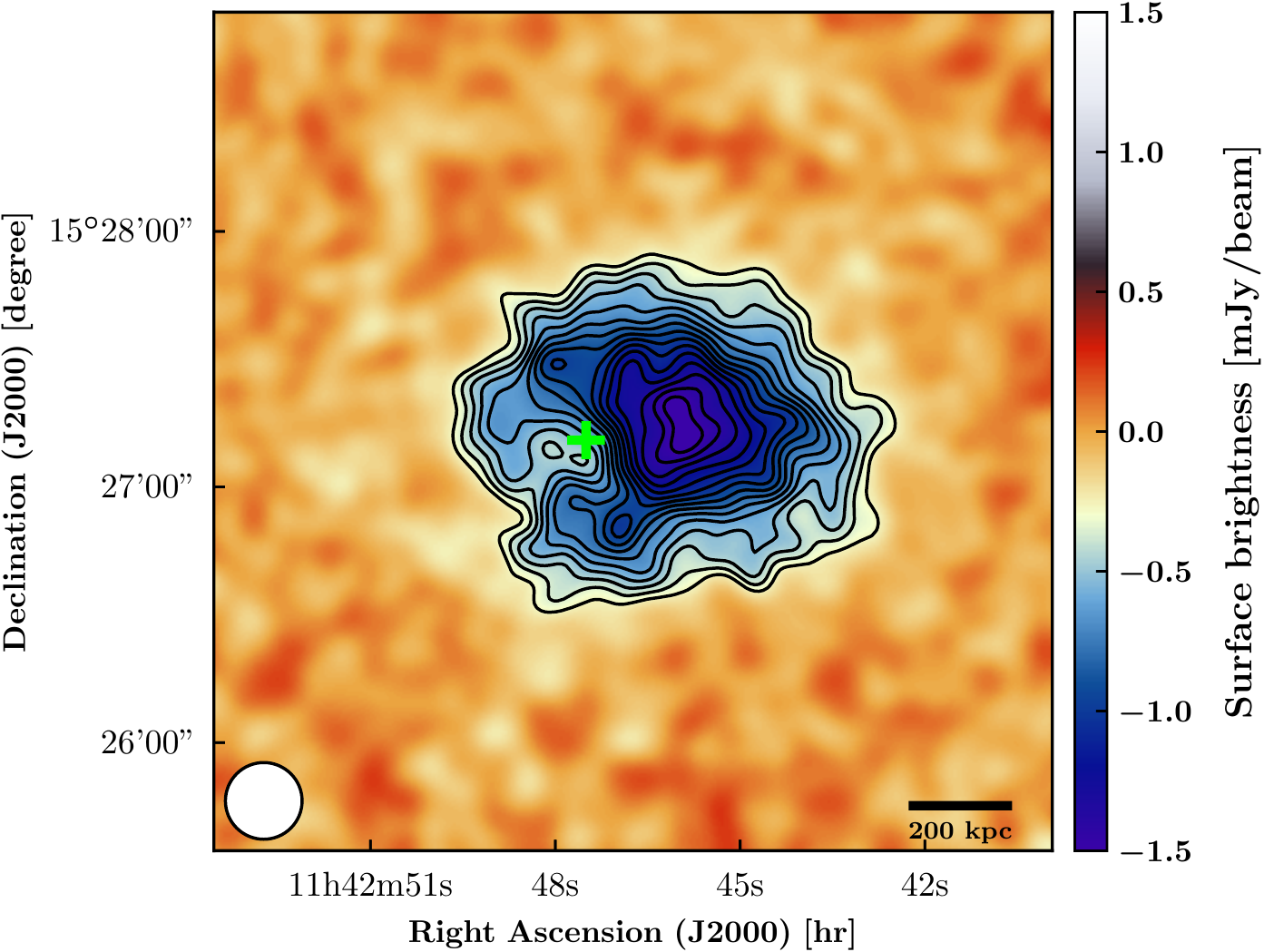}
\hspace{0.2cm}
\includegraphics[height=4.7cm]{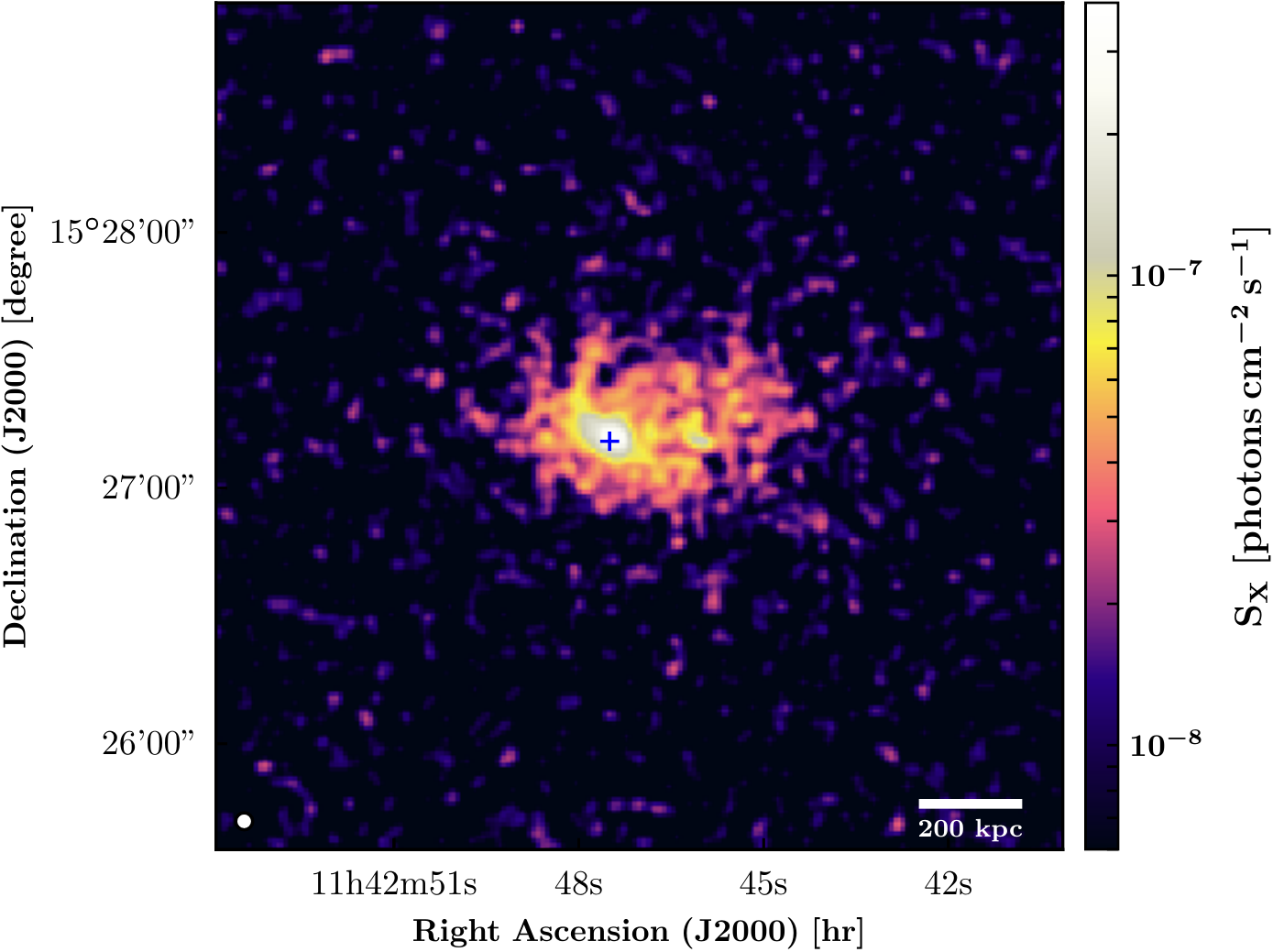}
\caption{{\footnotesize \textbf{Left:} NIKA2 surface brightness maps of \moo\ at 150 GHz smoothed with an additional 8~arcsec FWHM Gaussian filter. Significance contours starting at $3\sigma$ with $1\sigma$ steps are shown in black. The white disk in the bottom corner gives the width of the NIKA2 beam at 150~GHz. \textbf{Right:} \chandra\ exposure-corrected map of \moo\ in the 0.7-7.0~keV band smoothed with an additional 3~arcsec FWHM Gaussian filter. The green and blue crosses in both panels give the location of the radio source in the cluster. Figure extracted from \citep{rup19b}.}}
\label{fig:nk2_chandra_maps}
\end{figure*}

\subsection{NIKA2 SZ observations}\label{subsec:nika2}

The NIKA2 SZ observations of \moo\ have been realized in October 2017 for an effective observing time of 10.4 hours (OpenTime: 082-17, PI: F. Ruppin). The observations have been conducted under good weather conditions with a mean zenith opacity of 0.19 at 150~GHz and a stable atmosphere. We follow the baseline calibration procedure described in \cite{per19} and measure an absolute calibration uncertainty of 6\% at 150~GHz. We use the pre-processing method detailed in \cite{ada15} for the selection of valid detectors and the removal of  cosmic ray glitches and cryogenic vibrations from the raw data. The spatially correlated contaminants induced by both the atmosphere and the detector readout system have been removed from the data using an iterative procedure \cite{rup18}. Following the procedure described in \cite{ada16}, we estimate the residual noise power spectrum in the final map at 150~GHz based on null maps and find that it is well modeled by a simple white noise component. The circular transfer function resulting from the NIKA2 observations and data processing at 150~GHz has been computed using simulations as described in \cite{ada15}.\\
The left panel of Fig. \ref{fig:nk2_chandra_maps} shows the surface brightness map of \moo\ obtained at the end of the analysis of the NIKA2 data at 150~GHz. Significant negative signal is detected up to ${\sim}1$~arcmin away from the SZ peak. The cluster appears to be elliptical with an E-W orientation of the projected major axis. A point source is detected in the NIKA2 map at 260~GHz at $25$~arcsec to the east of the SZ peak at 150~GHz (green cross in Fig. \ref{fig:nk2_chandra_maps}). This position matches the one of a radio source found in the Faint Images of the Radio Sky at Twenty-Centimeters survey at 1.4~GHz \citep{bec95}. The negative SZ signal induced by \moo\ at 150~GHz is partly compensated by the signal emitted by this source. Measurements of the flux of this source have been obtained at 153~MHz by the TIFR GMRT Sky Survey \citep{int17} and at 31~GHz by the Combined Array for Research in Millimeter-wave Astronomy \citep[CARMA,][]{gon15}. We fit the spectral energy distribution of the source using a power law model based on all these measurements in order to estimate the flux of this contaminant at 150~GHz. We find a flux $F_{\rm{150~GHz}} = 1.5 \pm 0.3~\rm{mJy}$. This estimate is used in Sect. \ref{sec:ICM_1D} to define a Gaussian prior on the source emission at 150~GHz.

\subsection{Chandra X-ray observations}\label{subsec:chandra}

\begin{figure*}
\centering
\includegraphics[height=4.3cm]{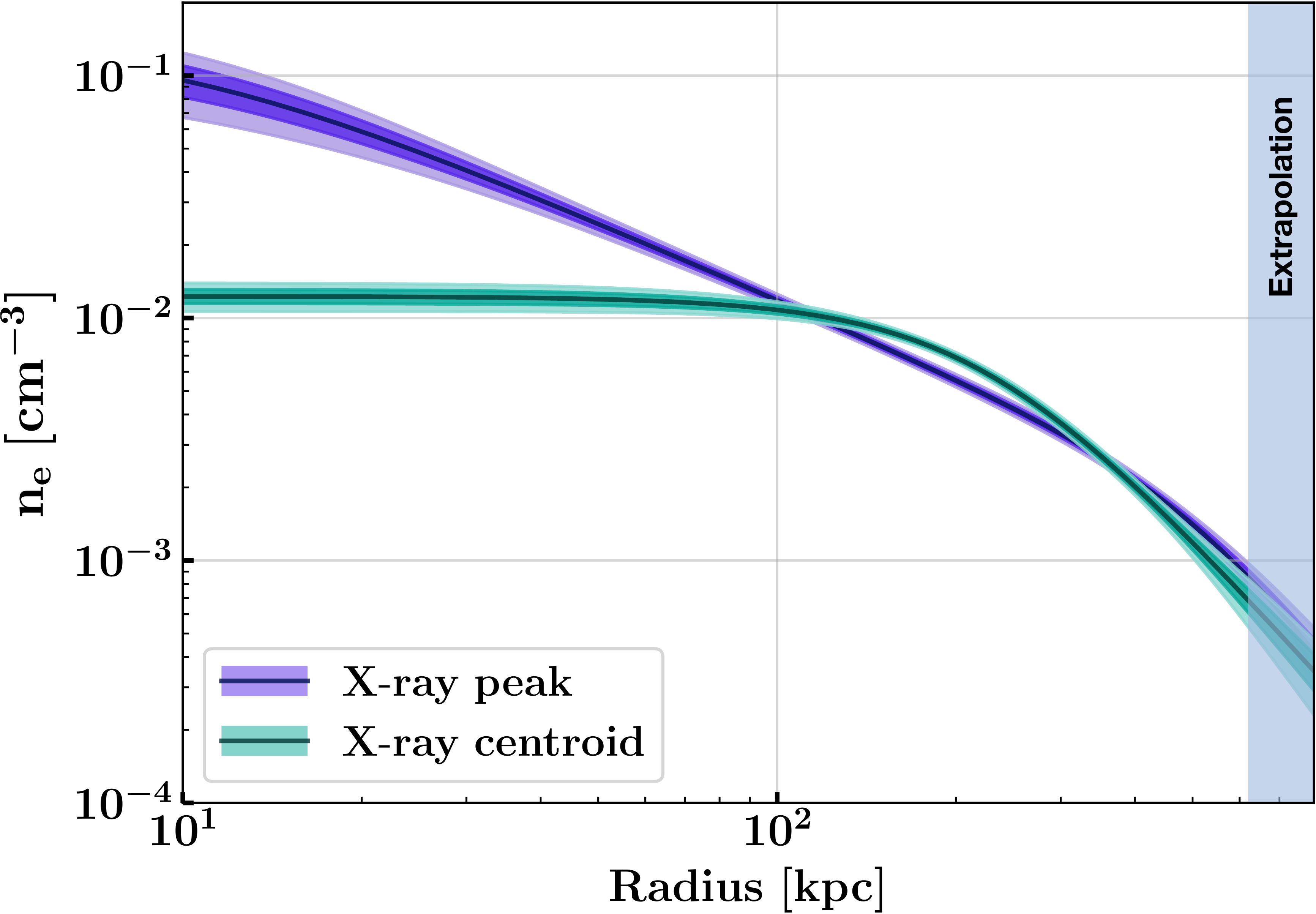}
\hspace{0.2cm}
\includegraphics[height=4.3cm]{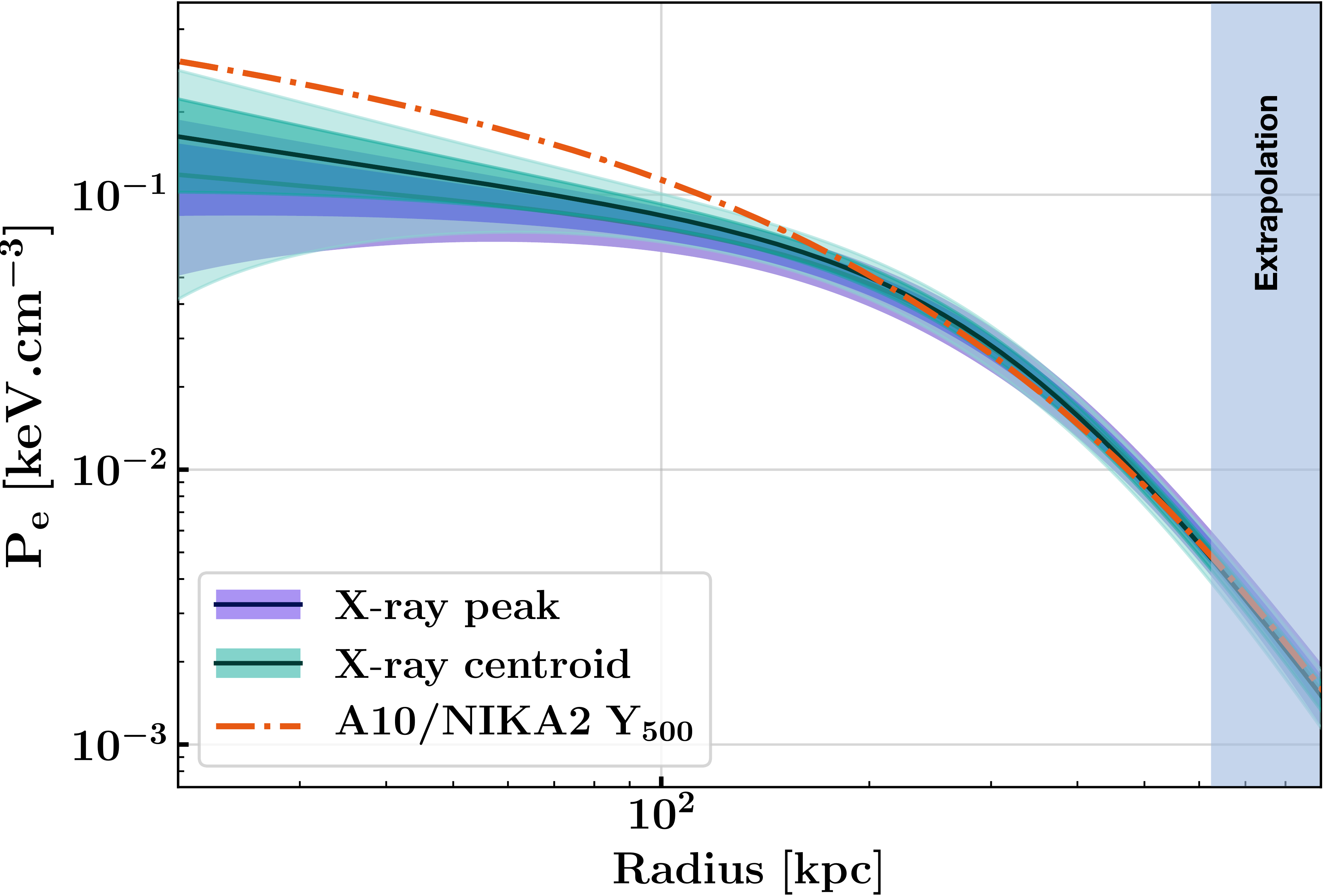}
\caption{{\footnotesize Density (left) and pressure (right) profiles of \moo\ estimated by considering the X-ray peak (purple) or the X-ray centroid (green) in the deprojection procedure based on the \chandra\ and NIKA2 data. The dark and light colored regions around the best-fit profiles give the 68\% and 95\% confidence levels. The light grey rectangles show the region where the models are extrapolated but not constrained by the data. Figure extracted from \citep{rup19b}.}}
\label{fig:Pe_ne_profiles}
\end{figure*}

The \chandra\ X-ray observations of \moo\ have been realized in February 2017 (ObsID: 18277, PI: S. A. Stanford) for a total cleaned exposure of 46.19~ks using the Advanced CCD Imaging Spectrometer I-chips. The data reduction is done using the \chandra\ Interactive Analysis of Observations software v4.10 based on the calibration database v4.8.0 provided by the \chandra\ X-ray Center. We follow the data reduction procedure detailed in \cite{mcd17} to reprocess the level 1 event files, remove flares from lightcurves, identify point sources, and estimate the X-ray background at the cluster location.\\
The right panel of Fig. \ref{fig:nk2_chandra_maps} shows the \chandra\ exposure-corrected surface brightness map of \moo\ after background and point source subtraction. Significant diffuse X-ray emission is detected within a region of similar angular size as the one obtained in SZ with NIKA2. The overall cluster morphology is similar to the one mapped by NIKA2 at 150~GHz. However, the X-ray peak (blue cross) is detected at ${\sim}100$~kpc from the SZ peak at the same location as the radio source found in the NIKA2 260~GHz map. The overpressure region to the west of the X-ray peak might result from an on-going merger event with a substructure (see \citep{rup19b} for more details about the dynamics of this cluster).\\
Given the disturbed dynamics of this cluster, we choose to consider both the X-ray centroid computed within a disk of 1~arcmin radius and the X-ray peak as deprojection centers. We extract an X-ray spectrum from the cleaned event list in the 0.7-7.0~keV band in a core-excised circular annulus with an external radius equal to $R_{500} = 790$~kpc. The latter is estimated iteratively using the  $M_{500}{-}T_X$ scaling relation from \cite{vik09}. We subtract the particle background from the spectrum and fit it jointly with the astrophysical background spectrum using two absorbed \texttt{APEC} \citep{smi01} models and a hard X-ray cosmic spectrum \texttt{BREMSS}. We find a mean ICM spectroscopic temperature of $T_X = 8.63 \pm 1.86~\mathrm{keV}$. We follow the methodology detailed in \cite{mcd17} to extract the cluster surface brightness profile $S_X$ in the 0.7-2.0~keV band in 20 annuli centered on both the X-ray peak and the X-ray centroid. This profile along with the spectroscopic temperature estimate are used to deproject the ICM density profile in Sect. \ref{sec:ICM_1D}.

\section{ICM thermodynamic properties from a multi-probe analysis}\label{sec:ICM_1D}

We fit the cluster emission measure profile $\mathrm{EM}(r) = \int n_e n_p \, dl = S_X / \Lambda(T_X,Z)$ using a Markov chain Monte Carlo (MCMC) analysis. The cooling function $\Lambda(T_X,Z)$ is obtained from the normalization of the X-ray spectrum assuming a constant metallicity $Z = 0.3 Z_{\odot}$. We assume a standard ionization fraction $n_e/n_p = 1.199$ \citep{and89} to estimate the proton density $n_p$. We model the electron density $n_e(r)$ using a a simplified Vikhlinin parametric model \citep[SVM;][]{vik06} based on five free parameters. The best-fit density models obtained at the end of the analysis of the two emission measure profiles centered on the X-ray peak and the X-ray centroid are shown in the left panel of Fig. \ref{fig:Pe_ne_profiles} in purple and green respectively. The density distributions are very well constrained from the cluster core up to $0.75 R_{500}$. The choice of deprojection center has a significant impact on the estimate of the ICM density distribution in the cluster core. This order of magnitude difference is caused by the peculiar morphology of the cluster hosting its highest density core at ${\sim}100$~kpc from its centroid.\\
\begin{figure*}
\centering
\includegraphics[height=4.6cm]{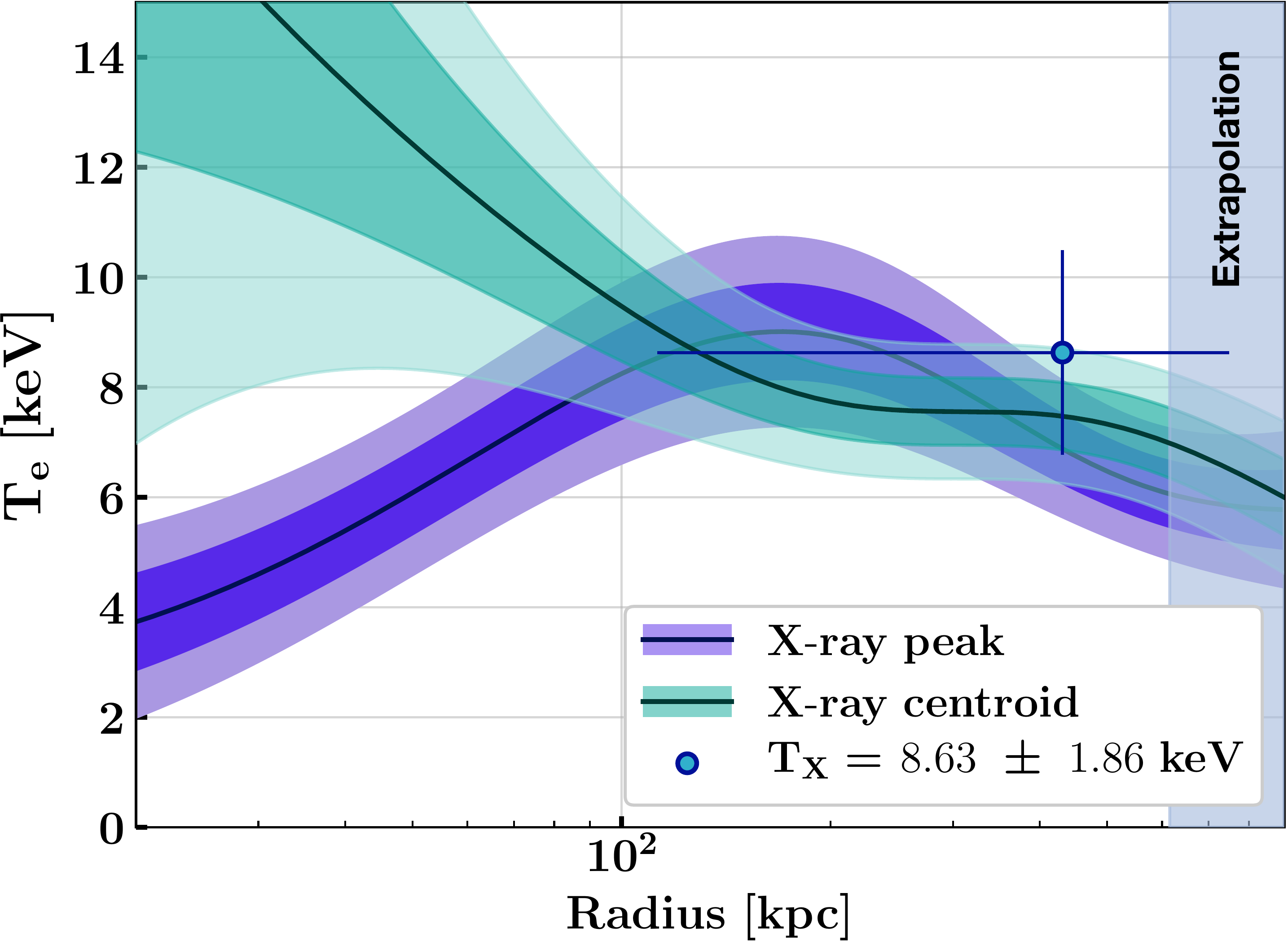}
\hspace{0.2cm}
\includegraphics[height=4.5cm]{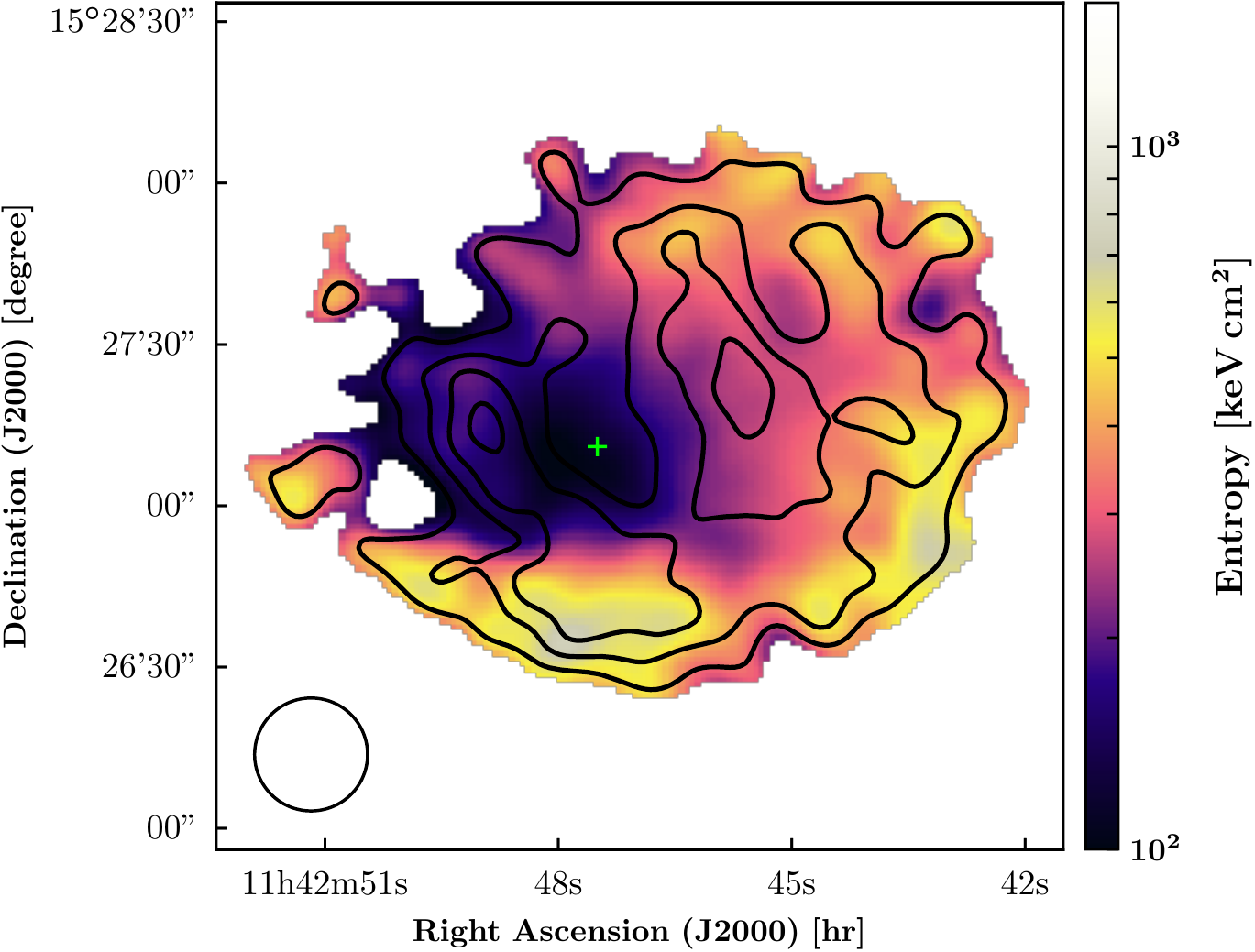}
\caption{{\footnotesize \textbf{Left:}  Temperature profiles obtained by combining the \chandra\ density profiles and the NIKA2 pressure profiles. We also show the single spectroscopic temperature measurement from the \chandra\ data. The color code is the same as in Fig. \ref{fig:Pe_ne_profiles}. \textbf{Right:} Entropy map of \moo\ obtained from the combination of the ICM pressure and density maps estimated using the NIKA2 and \chandra\ surface brightness maps shown in Fig. \ref{fig:nk2_chandra_maps}. The green cross shows the location of the X-ray peak. Figure extracted from \citep{rup19b}.}}
\label{fig:Te_profile_map}
\end{figure*}We estimate the  pressure profile of \moo\ using the pipeline developed for the NIKA2 SZ large program \citep{rup18}. The pressure distribution is modeled by a generalized Navarro-Frenk-White model \citep[gNFW;][]{nag07} characterized by five free parameters. Theses parameters are estimated using a MCMC analysis based on the NIKA2 SZ map at 150~GHz and the integrated Compton parameter measured by CARMA \citep{gon15}. We show the pressure profiles obtained by using the X-ray peak and the X-ray centroid as deprojection centers in the right panel of Fig. \ref{fig:Pe_ne_profiles}. Similarly to the \chandra\ density profile, very good constraints are obtained from the cluster core up to $0.75 R_{500}$. However, we do not measure any significant impact of the choice of deprojection center on the inner slope of the profile. This is due to the fact that the highest SZ signal is measured at a near constant amplitude over a region that encloses both the X-ray peak and the X-ray centroid. We also find that the inner slope of these profiles is slightly shallower than the one of the universal pressure profile from \cite{arn10} (orange dashdotted line) which is consistent with the disturbed cluster dynamics.\\
We estimate the other ICM profiles of this cluster by combining the \chandra\ density profile and the NIKA2 pressure profile. The ICM temperature profiles are shown in the left panel of Fig. \ref{fig:Te_profile_map}. They are computed under the ideal gas assumption such that $k_B T_e(r) = \frac{P_e(r)}{n_e(r)}$ where, $k_B$ is the Boltzmann constant. We find significant differences between the two profiles at radii $r \lesssim 100$~kpc. On the one hand, the profile estimated from the X-ray peak (purple) has a shape that is typical of a cool-core cluster with a low central value of ${\sim}4$~keV and a maximum of $9$~keV at $170$~kpc from the center. On the other hand, the profile obtained by using the X-ray centroid (green) is decreasing from a very high central temperature around 14~keV and reaches a plateau at ${\sim}7.5$~keV at $r\gtrsim 200$~kpc. Both profiles are compatible with the \chandra\ spectroscopic temperature estimate (blue point). We stress that reaching the same relative uncertainty on the temperature profile using X-ray spectroscopy would require at least an order of magnitude increase in exposure at this redshift.\\
We follow the method of \cite{ada17b} to estimate the maps of the ICM pressure ($\bar{P}_e$) and density ($\bar{n}_e$) based on the NIKA2 and \chandra\ surface brightness maps. We subtract the radio source signal from the NIKA2 map and deconvolve it from the analysis transfer function to minimize the bias induced by the filtering of the SZ signal. We process the \chandra\ point source and background subtracted surface brightness map after vignetting correction through an adaptative filter. This allows us to produce a density map at the same effective angular resolution as the NIKA2 pressure map. We combine these results to estimate the map of the ICM entropy $\bar{K}_e = \bar{P}_e / \bar{n}_e^{5/3}$ (see right panel in Fig. \ref{fig:Te_profile_map}). We find that the entropy measured at the location of the X-ray peak is three times lower than the mean entropy value measured in the whole map. The core entropy is enclosed between 100 and $200~\mathrm{keV.cm^2}$ within a region extending over a ${\sim}120$~kpc radius from the X-ray peak. This supports the presence of a cool-core which is not disturbed by the merging event with the infalling substructure yet.

\section{Conclusions}\label{sec:conclusions}

We have realized a state-of-the-art characterization of the properties of the massive cluster \moo. We have jointly analyzed the spatially resolved NIKA2 SZ and \chandra\ X-ray observations of this cluster to estimate profiles and maps of its ICM thermodynamic properties. This allows us to conclude that this cluster is an on-going merger hosting a cool-core at the same location as a radio source identified in previous surveys. This is the first time that such a detailed ICM analysis is performed at $z>1$ to the best of our knowledge.

\section*{Acknowledgements}
\small{We would like to thank the IRAM staff for their support during the campaigns. The NIKA dilution cryostat has been designed and built at the Institut N\'eel. In particular, we acknowledge the crucial contribution of the Cryogenics Group, and in particular Gregory Garde, Henri Rodenas, Jean Paul Leggeri, Philippe Camus. This work has been partially funded by the Foundation Nanoscience Grenoble and the LabEx FOCUS ANR-11-LABX-0013. This work is supported by the French National Research Agency under the contracts "MKIDS", "NIKA" and ANR-15-CE31-0017 and in the framework of the "Investissements d’avenir" program (ANR-15-IDEX-02). This work has benefited from the support of the European Research Council Advanced Grant ORISTARS under the European Union's Seventh Framework Programme (Grant Agreement no. 291294). We acknowledge fundings from the ENIGMASS French LabEx (R. A. and F. R.), the CNES post-doctoral fellowship program (R. A.), the CNES doctoral fellowship program (A. R.) and the FOCUS French LabEx doctoral fellowship program (A. R.). R.A. acknowledges support from Spanish Ministerio de Econom\'ia and Competitividad (MINECO) through grant number AYA2015-66211-C2-2. This work has benefited from the support of the European Research Council Advanced Grants ORISTARS and M2C under the European Unions Seventh Framework Programme (Grant Agreement Nos. 291294 and 340519). Support for this work was provided by NASA through SAO Award Number SV2-82023 issued by the CXC, which is operated by the Smithsonian Astrophysical Observatory for and on behalf of NASA under contract NAS8-03060.}

%
%
%

\end{document}